\documentclass[prd,showpacs,preprintnumbers]{revtex4} % Physical Review D
\usepackage{amsmath}    % need for subequations
\usepackage{graphicx}   % need for figures
\usepackage{verbatim}   % useful for program listings
\usepackage{color}      % use if color is used in text
\usepackage{subfigure}  % use for side-by-side figures
\usepackage{hyperref}   % use for hypertext links, including those to external documents and URLs

\begin{document}
\bibliographystyle{plainnat}

\preprint{KOBE-TH-15-13}
\title{No-Go Theorem for Gauss-Bonnet Inflation without Inflaton Potential}
\author{Getbogi Hikmawan$^{1,2}$, Jiro Soda$^{2}$, Agus Suroso$^{1}$, and Freddy P. Zen$^{1}$}

\affiliation{${}^{1)}$Department of Physics, Institut Teknologi Bandung, Bandung 40132, Indonesia \\
${}^{2)}$Department of Physics, Kobe University, Kobe 657-8501, Japan}
\date{\today}

\begin{abstract}
Recently, an interesting inflationary scenario, named Gauss-Bonnet inflation, 
is proposed by Kanti et al.~\cite{Kanti:2015pda,Kanti:2015dra}. In the model, there is no inflaton potential but 
the inflaton couples to the Guass-Bonnet term. In the case of quadratic coupling, they find inflation occurs with graceful exit. 
The scenario is attractive because of the natural set-up. However, we show there exists the gradient instability in the tensor perturbations in this inflationary model. We further prove the no-go theorem for the Gauss-Bonnet inflation without an inflaton potential.  
\end{abstract}
\pacs{98.80.Cq, 04.50.Kd}

\maketitle

%\tableofcontents

\section{Introduction}

It is believed that the most promising candidate for the ultimate unified theory 
is superstring theory.
The effective action stemming from superstring theory contains 
higher order curvature terms.
Indeed, in four dimensions, the Gauss-Bonnet (GB) term appears as a one-loop string correction~\cite{Antoniadis:1992sa}. Motivated by this fact,
Einstein-scalar-Gauss-Bonnet theory is studied and non-singular cosmological solutions are found~\cite{Antoniadis:1993jc,Easther:1996yd,Kawai:1998bn,Kanti:1998jd}.
The solutions have super inflation phase where the Hubble parameter increases, 
which indicates  the violation of the weak energy condition. 
Subsequently, cosmological perturbations in this background are investigated~\cite{Kawai:1997mf,Kawai:1998ab,Soda:1998tr,Hwang:1999gf,
Kawai:1999pw,Kawai:1999xn,Cartier:2001is}.
In the process,   the so-called gradient instability is  found for the first time in \cite{Kawai:1998ab,Soda:1998tr}. 
Nowadays, this is known as a useful criterion for model selections in various extension of general relativity~\cite{Kobayashi:2011nu}.
By applying this criterion, it has been shown the non-singular solutions are unstable
in general because of the violation of the weak energy condition~\cite{Kawai:1997mf,Kawai:1998ab,Soda:1998tr,
Kawai:1999pw,Kawai:1999xn}.

Recently, Kanti et al.~\cite{Kanti:2015pda,Kanti:2015dra} analyzed the Einstein-scalar-Gauss-Bonnet theory with a quadratic coupling function
 and found that the theory contains inflationary solutions 
where de Sitter phase possesses a natural exit mechanism
and are replaced by linearly expanding Milne phases. Remarkably, there is no inflaton potential in this model. 
They also claimed that only the quadratic coupling leads to these results among monomial coupling functions. In this sense, the model is simple and unique.
Hence, it is worth investigating the inflationary scenario, named the Gauss-Bonnet inflation.

In order for this GB inflation to be viable, its predictions must be compatible with 
current observational data~\cite{Planck:2013jfk}. Hence, we need to calculate perturbations and compare its predictions with observations. 
First, we calculate tensor perturbations in the inflationary background following previous results~\cite{Kawai:1998ab,Soda:1998tr}. 
Unfortunately, we find the gradient  instability in tensor perturbations 
 in the GB inflation. Hence, the model is not viable phenomenologically.
We also extend this result to more
general coupling functions and establish the no-go theorem for GB inflation
without an inflaton potential.

The organization of the paper is as follows. In section II, we review the cosmological inflationary solutions obtained by Kanti et al.~\cite{Kanti:2015pda,Kanti:2015dra}.
In section III, we numerically examine the dynamics of tensor perturbations and find the instability of GB inflation. In section IV, we prove a no-go theorem for GB inflation without an inflaton potential. As a byproduct, we also present the stability condition for 
GB inflation with an inflaton potential.
The final section is devoted to conclusion. 

\section{Review of Gauss-Bonnet Inflation}
In this section, we review the main results of the papers ~\cite{Kanti:2015pda,Kanti:2015dra}. In particular, we show there exists
a quasi-de Sitter phase in the Einstein-scalar-Gauss-Bonnet theory with a quadratic 
coupling function
in spite of absence of an inflaton potential.  

The action considered there is GB gravity with a scalar field $\phi$ which
 coupled non-minimally with a coupling function $f(\phi)$ to gravity via the GB term $ R_{GB}^{2} $
\begin{equation}
\label{eq:ex1}
\mathit{S}=\int d^4x \sqrt{-g}\biggl[\frac{R}{2\kappa^{2}}-\frac{1}{2}(\nabla\phi)^{2}+\frac{1}{8} f(\phi) R_{GB}^{2}\biggr].
\end{equation}
where $\kappa^{2} \equiv 8\pi G $ is the gravitational coupling constant, $g$ is a determinant of the metric $g_{\mu\nu}$,
$R$ is the Ricci scalar,  and the GB term is given by
\begin{eqnarray}
R_{GB}^{2} = R^{\mu\nu\rho\lambda} R_{\mu\nu\lambda\rho}  - 4 R^{\mu\nu} R_{\mu\nu} +R^2  \ .
\end{eqnarray}
Note that there is no inflaton potential $V(\phi)$.
The variation of this action with respect to the metric tensor and scalar field give us the field equations,
\begin{equation}
R_{\mu\nu}-\frac{1}{2}g_{\mu\nu}R+P_{\mu\alpha\nu\beta}\nabla^{\alpha\beta}f
=\partial_{\mu}\phi\partial_{\nu}\phi - \frac{1}{2} g_{\mu\nu}(\nabla\phi)^2,	
\end{equation}
and
\begin{equation}
\frac{1}{\sqrt{-g}}\partial_{\mu}\left[ \sqrt{-g}\partial^{\mu}\phi \right]+\frac{1}{8}  f' R_{GB}^2 =0 \ ,
\end{equation}	
where we defined $ f'\equiv df/d\phi$, $\kappa^{2} $ is set to unity and $P_{\mu\alpha\nu\beta}$ is defined as
\begin{equation}
P_{\mu\alpha\nu\beta}= R_{\mu\alpha\nu\beta}+2g_{\mu[\beta}R_{\nu]\alpha}+2g_{\alpha[\nu}R_{\beta]\mu}+Rg_{\mu[\nu}g_{\beta]\alpha}.
\end{equation}

For a homogeneous and isotropic flat spacetime, the metric is described by a scale factor $a(t)$ as
\begin{equation}
ds^2=-dt^2+a^2(t)\ \delta_{ij}dx^{i}dx^{j} \ . 
\end{equation}
Now, the explicit form of the field equations reads
\begin{eqnarray}
\label{2}
6H^{2}(1+H\dot{f})=\dot{\phi^{2}}  \ , \\
\label{3}
2(1+H\dot{f})(H^{2}+\dot{H})+H^{2}(1+\ddot{f})=-\frac{1}{2} \dot{\phi^{2}} \ ,  \\
\label{4}
\ddot{\phi}+3H\dot{\phi}-3f'H^{2}(H^{2}+\dot{H})=0  \ ,
\end{eqnarray}
where $H\equiv \dot{a}/a$ is the Hubble parameter and a dot denotes a derivative with respect to time.

To be concrete, we need to specify the function $f(\phi)$. 
Kanti et al. ~\cite{Kanti:2015pda,Kanti:2015dra} found the quadratic coupling function $  f=\lambda\phi^{2} $ with a coupling constant $\lambda$ 
 gives rise to the inflationary solutions. In the early time, that is, in the strong gravity regime,  the GB term dominates over the Ricci scalar term. 
Indeed, the dynamics of the universe  does not show any significant change  
as long as the scalar field takes very large values or  $\lambda$
takes a very large value
even if we neglect the Ricci term. 
Therefore, we will neglect the Ricci term for a moment.
With this approximation, the unity terms inside of brackets in equation (\ref{2}) and (\ref{3}) can be neglected. 
Thus,  we obtain the equation
\begin{equation}
\label{5}
\dot{H}  + H^2 \left( 1 - \frac{H^2}{H_{dS}^2} \right)  =0 \ . 
\end{equation}
where $H_{dS}^2 = - 5/24\lambda$. From now on, we assume $\lambda <0$.

Apparently, there exists a solution with $\dot{H}=0$. 
For this case, Eq.(\ref{5}) can be integrated as
\begin{equation}
\label{6a}
a(t)=a_0 \exp \left( H_{dS} \ t \right) \ .	
\end{equation}
From the solution (\ref{6a}), we clearly see inflation can happen due to the GB term. 
Moreover, the scalar field can be solved as
\begin{equation}
\label{6b}
\phi=\phi_0 \exp\left(-\frac{5}{2} \ H_{dS} \ t \right) \ .
\end{equation}

\begin{figure}
\centering
\includegraphics[width=15cm]{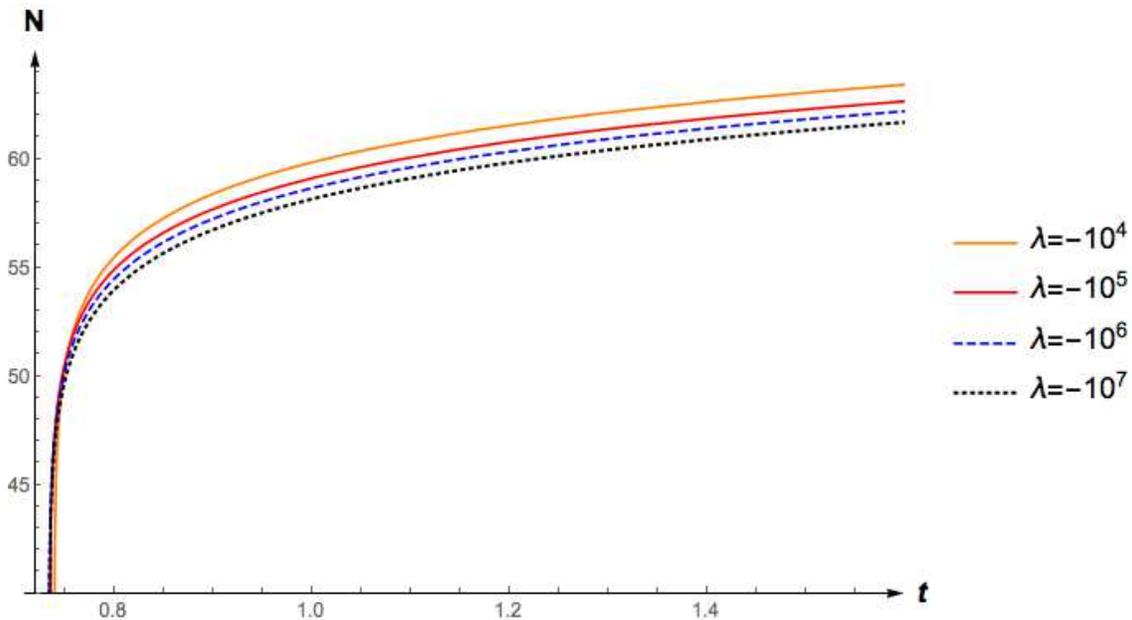}
\caption{The number of e-foldings, $N=Log(a)$ is plotted as a function of time for various $\lambda<0$.}
\label{fig:egb}
\end{figure}

%\begin{figure}[h]
%\centering
%\includegraphics[width=15cm]{"Plot of Einstein-scalar-Gauss-Bonnet Theory(2)"}
%\caption{The number of e-foldings, $N=Log(a)$ is plotted as a function of time for various $\lambda<0$.}
%\label{fig:PlotofEinstein-scalar-Gauss-BonnetTheory(2)}
%\end{figure}

Even for general cases, we can analytically integrate Eq.(\ref{5}) as
\begin{eqnarray}
   H = \frac{H_{dS}}{\sqrt{1 + \frac{2C}{5} H_{dS}^2 a^2 }}  \ ,
\label{H_sol}
\end{eqnarray}
where $C$ is a constant of integration. As you can see, as the universe expands, the Hubble parameter decreases. Hence, we have to tune $\nu^{-2} \equiv 2C H_{dS}^2 / 5
=C/12\lambda $ so that inflation  occurs for a sufficiently long time.
In this case, the scale factor  can be obtained by analytically solving 
Eq.(\ref{H_sol}) using the change of a variable $a=\nu \tan \omega$. The resultant solution is given by
\begin{equation}
\label{6c}
\sqrt{a^2+\nu^2}+\nu \log (\frac{\sqrt{a^2+\nu^2}-\nu}{a})=\sqrt{\frac{5}{2C}}(t+t_0) \ ,
\end{equation}
where $t_0$ is a constant of integration.

Because of the implicit solution for the scale factor $a$ in equation (\ref{6c}), the explicit solution for scalar field can not be found explicitly as a function of time.
Yet, we can obtain it as a function of the scale factor,
\begin{equation}
\label{6d}
\phi=C_0\bigl(\frac{2C}{5}\bigr)^{5/4}\frac{(a^2+\nu^2)^{5/4}}{a^{5/2}} \ ,
\end{equation}
where $C_0$ is another constant of integration.
From this equation, we can see that the scalar field initially decreases. 
In the regime $a^2\gg\nu^2$, however, the scalar field becomes constant.

We have also numerically solved  Eqs.(\ref{2}) ,(\ref{3}) , and (\ref{4}) without any approximation.
The time evolution of the scale factor in the term of the number of e-foldings, $N=log(a)$ is plotted in Fig.(\ref{fig:PlotofEinstein-scalar-Gauss-BonnetTheory(2)}).
Here you can see inflationary behavior and the subsequent decelerating expansion
of the universe. 
Notice that 
the initial inflationary period can be elongated by making $\nu$ (or $\lambda$) large.

\section{Instability of quadratic coupling models}

In the previous section, we have shown that there exists inflation without an inflaton potential.
Given the inflationary background, the next step is to calculate perturbations and check if the model is 
compatible with observational data. Unfortunately, however, we will see the instability of the Gauss-Bonnet inflation.

Here, we begin with tensor perturbations defined by 
\begin{equation}
\label{6}
ds^{2}=-dt^2+a^2(\delta_{ij}+h_{ij}dx^{i}dx^{j})
\end{equation}
where $h_{,j}^{ij}=h_i^i=0$. For the metric (\ref{6}), we can obtain the quadratic action for  the tensor perturbations as follows,
\begin{equation}
\label{7}
\mathit{S}=\frac{1}{8}\int d^4xa^3[(1+H\dot{f})\dot{h_{ij}}\dot{h^{ij}}-\frac{1}{a^2}(1+\ddot{f})h_{ij,k}h^{ij,k}].
\end{equation}

\begin{figure}
\centering
\includegraphics[width=10cm]{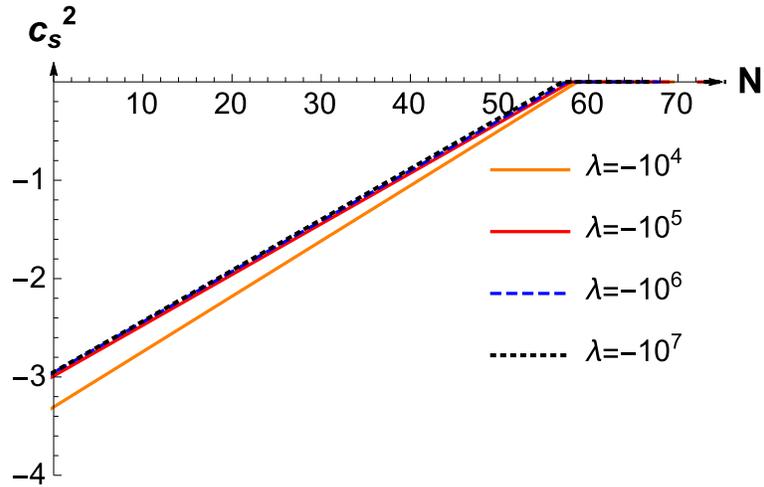}
\caption{The time evolution of the squared of the effective speed of sound $c_s^2$ of the model is plotted  as a function of the number of e-foldings.}
\label{fig:plotstability}
\end{figure}

%
%\begin{figure}
%\centering
%\includegraphics[width=12cm]{"plot stability"}
%\caption{The time evolution of the squared of the effective speed of sound $c_s^2$ of the model is plotted  as a function of the number of e-foldings.}
%\label{fig:plotstability}
%\end{figure}
%
Therefore, the equation for tensor perturbations can be found as follows,
\begin{equation}
\label{acten}
\ddot{h_{ij}}+(3H+\frac{\dot{\alpha}}{\alpha})\dot{h_{ij}}+\frac{k^2}{a^2}\frac{1+\ddot{f}}{\alpha} h_{ij}=0,
\end{equation}
where $\alpha=(1+H\dot{f})$ is defined and $k$ is a wavenumber. 
Note that the system contains ghost if $\alpha <0$. Since the existence of ghost 
implies that the model is pathological, we must avoid this possibility.
From Eq. (\ref{acten}), we can see that the stability of the system is determined by the last term, while the second term is a friction term. The model is stable for tensor perturbation when squared the effective speed of sound is positive
\begin{equation}
\label{9}
c_s^2 \equiv \frac{1+ \ddot{f}}{\alpha} >0 \ .
\end{equation}
Therefore, we can check the stability of tensor perturbations of the model by checking
 the value of $c_s^2$. From the equations of motion (\ref{2}) to (\ref{4}),
 we can get the time evolution for $c_s^2$ for several $\lambda$ values
 as can be seen in Fig.\ref{fig:plotstability}. Because $c_s^2$ is negative during  inflation,  we can conclude that the system is unstable under tensor perturbations.

In order to see the instability of the system more explicitly, we numerically solved the time evolution of the tensor perturbations.
\begin{figure}
\centering
\includegraphics[width=15cm]{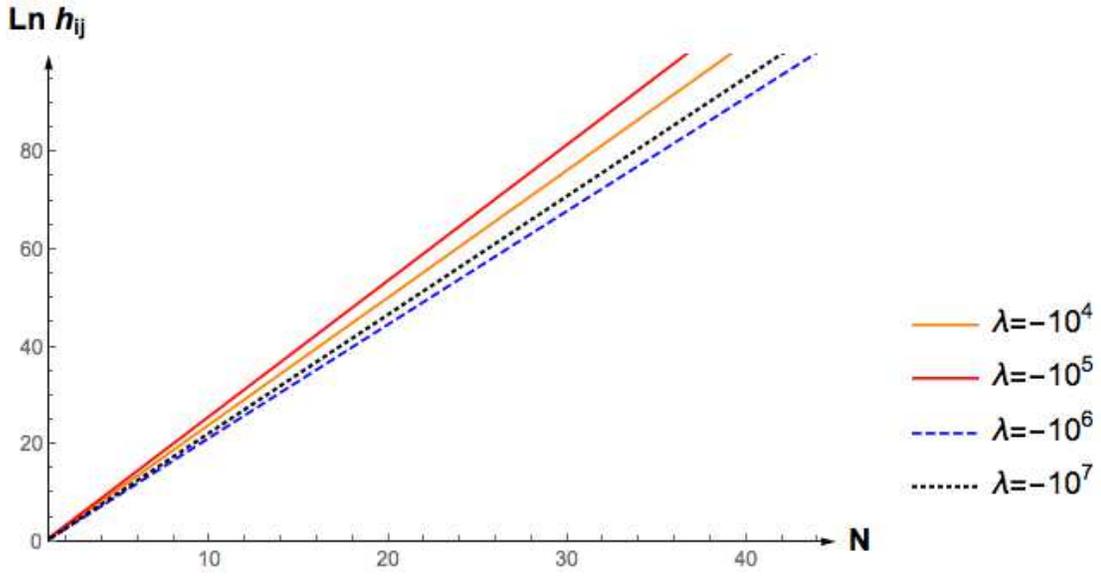}
\caption{The time evolution of the tensor perturbations $h_{ij}$ is plotted  as a function of the number of e-foldings. Here, we put $k=1$.}
\label{fig:EvolutionofTensorPerturbationModes}
\end{figure}

%\begin{figure}
%\centering
%\includegraphics[width=15cm]{"Evolution of Tensor Perturbation Modes"}
%\caption{The time evolution of the tensor perturbations $h_{ij}$ is plotted  as a function of the number of e-foldings. Here, we put $k=1$.}
%\label{fig:EvolutionofTensorPerturbationModes}
%\end{figure}
From Fig.\ref{fig:EvolutionofTensorPerturbationModes}, we can see that the tensor perturbations grow rapidly. Therefore, the system is unstable under tensor perturbations.

\section{No-Go Theorem for Gauss-Bonnet Inflation}
In the previous section, we have shown that the instability of GB theory for the specific coupling function
$f(\phi) = \lambda \phi^2 $. In this section, we will extend this result to more general coupling functions and prove a no-go theorem for GB inflation without an inflaton potential.

We shall start with the action 
\begin{equation}
\label{10}
\mathit{S}=\int d^4x \sqrt{-g}\biggl[\frac{R}{2\kappa^{2}}-\frac{1}{2}(\nabla\phi)^{2}-V(\phi)+\frac{1}{8}f(\phi) R_{GB}^{2}\biggr] \ ,
\end{equation} 
where we have incorporated an inflaton potential into the model.
The field equations for this action are similar to Eqs.(\ref{2})-(\ref{4}).
The Hamiltonian constraint reads
\begin{equation}
\label{11}
6H^{2}(1+H\dot{f})=\dot{\phi^{2}} +2V(\phi) \ ,
\end{equation}
and the Einstein equation gives us the following evolution equations
\begin{equation}
\label{12}
2(1+H\dot{f})(H^{2}+\dot{H})+H^{2}(1+\ddot{f})=-\frac{\dot{\phi^{2}}}{2}+V(\phi) \ ,
\end{equation}
Finally, we obtain the scalar field equation
\begin{equation}
\label{13}
\ddot{\phi}+3H\dot{\phi}-3f'H^{2}(H^{2}+\dot{H})-\frac{dV}{d\phi}=0 \ .
\end{equation}
From the background field equations (\ref{11}) and (\ref{12}), we get a relation
\begin{equation}
\begin{aligned}
	(1+ \ddot{f})&=\frac{2V}{H^2}-(1+H\dot{f})(\frac{2\dot{H}}{H^2}+5)\\
	&=\frac{2V}{H^2}+\alpha(2 \epsilon_H -5) \ ,
\end{aligned}
\end{equation}
where we used $\alpha$ defined before and the slow roll parameter $\epsilon_H = - \dot{H}/H^2$.
Therefore, the final expression of the action can be written as follows
\begin{equation}
\mathit{S}=\frac{1}{8}\int d^4x a^3\alpha \left[ 
             \dot{h}_{ij}\dot{h}^{ij}-\frac{1}{a^2}\left\{ (2\epsilon_H -5)a + \frac{2V}{H^2\alpha}\right\} h_{ij,k}h^{ij,k}\right],
\end{equation}
and the Hamiltonian is given by
\begin{equation}
	\label{14}
	\mathit{H}=\int d^4x[\frac{2\pi^{ij}\pi_{ij}}{a^3\alpha}+[\frac{Va}{4H^2}+\frac{1}{8}(2\epsilon_H-5)a\alpha] h_{ij,k}h^{ij,k}].
\end{equation}
From the Hamiltonian in  Eq.(\ref{14}), we can see that the model will be stable if $\alpha >0$ and
\begin{equation}
\Biggl[\frac{V}{4H^2}+\frac{1}{8}(2\epsilon_H -5)\alpha\Biggr]>0 \ .
\label{main}
\end{equation}

Now, let us consider models without an inflaton potential, namely $V=0$. 
Suppose that quasi-de Sitter inflation occurs for some coupling function $f(\phi)$.
Because of the quasi-de Sitter inflation, the slow roll parameter satisfies the condition $\epsilon_H \ll 1$. Hence, we always have the inequality  $2\epsilon_H -5<0$.
 Thus, the condition (\ref{main}) is satisfied only for the models $\alpha<0$. However, that implies the existence of the ghost. Namely, inflationary solutions must be unstable or contain ghost. In any case, the model is not phenomenologically allowed.
 Thus, we proved a no-go theorem for GB inflation 
without an inflaton potential.
Therefore, the cosmological model with the GB term without inflaton potential can not be a viable model. 

Even when an inflaton potential, $V$, is included,  the stability condition (\ref{main}) still remains useful. Indeed, it can be used to select viable models. 
In fact, because of the variety of the form of potential function, it is possible to find  the stable cosmological model with the GB term.

\section{conclusion}

We studied Einstein-scalar-Gauss-Bonnet theory with non-minimal coupling function. 
In particular, we focussed on inflation without an inflaton potential. 
In this model, in the case of quadratic coupling,  inflation occurs and possess a natural exit mechanism. The scenario is attractive because of the simple set-up. 
This motivated us to examine perturbative stability of the inflationary background 
solutions. We numerically solved the dynamics of tensor perturbations and
 found the gradient instability~\cite{Soda:1998tr} in  tensor perturbations in the inflationary model. 
We further extended this result and proved the no-go theorem for the GB inflation without an inflaton potential.  Thus, we have shown that the GB inflation without an inflaton potential is not phenomenologically viable.

Of course, if we incorporate an inflaton potential into the model, there are stable inflationary solutions~\cite{Neupane:2007qw}. 
We have also given the stability  criterion for this class of models. 
For stable models, we can discuss phenomenological predictions and compare them 
with observational data.
In fact, at high energy, it is natural to consider Gauss-Bonnet term as a correction~\cite{Weinberg:2008hq}.
Hence, it is still intriguing to study Einstein-scalar-Gauss-Bonnet theory in the cosmological context.

\begin{acknowledgments}
This work was supported by  Grants-in-Aid for Scientific Research (C) No.25400251
 and "MEXT Grant-in-Aid for Scientific Research on Innovative Areas  No.26104708 and “Cosmic Acceleration”(No.15H05895) . Part of this work by G. H., F. P. Z., and A. S. is supported by "Riset Inovasi KK ITB 2015" (No. 237j/I1.C01/PL/2015), "Riset Desentralisasi ITB 2015" (No. 310e/I1.C01/PL/2015 and No. 311i/I1.C01/PL/2015), "Riset PMDSU 2015"　(No. 314h/I1.C01/PL/2015), and "PKPI Scholarship" from Ministry of Research, Technology, and Higher Education of the Republic of indonesia.
\end{acknowledgments}


\begin{thebibliography}{50}

%\cite{Kanti:2015pda}
\bibitem{Kanti:2015pda} 
  P.~Kanti, R.~Gannouji and N.~Dadhich,
  %``Gauss-Bonnet Inflation,''
  Phys.\ Rev.\ D {\bf 92}, no. 4, 041302 (2015)
  [arXiv:1503.01579 [hep-th]].
  %%CITATION = ARXIV:1503.01579;%%

%\cite{Kanti:2015dra}
\bibitem{Kanti:2015dra} 
  P.~Kanti, R.~Gannouji and N.~Dadhich,
  %``Early-time cosmological solutions in Einstein-scalar-Gauss-Bonnet theory,''
  Phys.\ Rev.\ D {\bf 92}, no. 8, 083524 (2015)
  [arXiv:1506.04667 [hep-th]].
  %%CITATION = ARXIV:1506.04667;%%

%\cite{Antoniadis:1992sa}
\bibitem{Antoniadis:1992sa} 
  I.~Antoniadis, E.~Gava and K.~S.~Narain,
  %``Moduli corrections to gravitational couplings from string loops,''
  Phys.\ Lett.\ B {\bf 283}, 209 (1992)
  [hep-th/9203071];\\
  %%CITATION = HEP-TH/9203071;%%
  I.~Antoniadis, E.~Gava and K.~S.~Narain,
  %``Moduli corrections to gauge and gravitational couplings in four-dimensional superstrings,''
  Nucl.\ Phys.\ B {\bf 383}, 93 (1992)
  [hep-th/9204030].
  %%CITATION = HEP-TH/9204030;%%

%\cite{Antoniadis:1993jc}
\bibitem{Antoniadis:1993jc} 
  I.~Antoniadis, J.~Rizos and K.~Tamvakis,
  %``Singularity - free cosmological solutions of the superstring effective action,''
  Nucl.\ Phys.\ B {\bf 415}, 497 (1994)
  [hep-th/9305025];\\
  %%CITATION = HEP-TH/9305025;%%
  J.~Rizos and K.~Tamvakis,
  %``On the existence of singularity free solutions in quadratic gravity,''
  Phys.\ Lett.\ B {\bf 326}, 57 (1994)
  doi:10.1016/0370-2693(94)91192-4
  [gr-qc/9401023].
  %%CITATION = doi:10.1016/0370-2693(94)91192-4;%%

%\cite{Easther:1996yd}
\bibitem{Easther:1996yd} 
  R.~Easther and K.~i.~Maeda,
  %``One loop superstring cosmology and the nonsingular universe,''
  Phys.\ Rev.\ D {\bf 54}, 7252 (1996)
  [hep-th/9605173].
  %%CITATION = HEP-TH/9605173;%%

%\cite{Kawai:1998bn}
\bibitem{Kawai:1998bn} 
  S.~Kawai and J.~Soda,
  %``Nonsingular Bianchi type 1 cosmological solutions from 1 loop superstring effective action,''
  Phys.\ Rev.\ D {\bf 59}, 063506 (1999)
  [gr-qc/9807060].
  %%CITATION = GR-QC/9807060;%%

%\cite{Kanti:1998jd}
\bibitem{Kanti:1998jd} 
  P.~Kanti, J.~Rizos and K.~Tamvakis,
  %``Singularity free cosmological solutions in quadratic gravity,''
  Phys.\ Rev.\ D {\bf 59}, 083512 (1999)
  [gr-qc/9806085].
  %%CITATION = GR-QC/9806085;%%

%\cite{Kawai:1997mf}
\bibitem{Kawai:1997mf} 
  S.~Kawai, M.~a.~Sakagami and J.~Soda,
  %``Perturbative analysis of nonsingular cosmological model,''
  gr-qc/9901065.
  %%CITATION = GR-QC/9901065;%%

%\cite{Kawai:1998ab}
\bibitem{Kawai:1998ab} 
  S.~Kawai, M.~a.~Sakagami and J.~Soda,
  %``Instability of one loop superstring cosmology,''
  Phys.\ Lett.\ B {\bf 437}, 284 (1998)
  [gr-qc/9802033].
  %%CITATION = GR-QC/9802033;%%

%\cite{Soda:1998tr}
\bibitem{Soda:1998tr} 
  J.~Soda, M.~a.~Sakagami and S.~Kawai,
  %``Novel instability in superstring cosmology,''
  gr-qc/9807056.
  %%CITATION = GR-QC/9807056;%%

%\cite{Hwang:1999gf}
\bibitem{Hwang:1999gf} 
  J.~c.~Hwang and H.~Noh,
  %``Conserved cosmological structures in the one loop superstring effective action,''
  Phys.\ Rev.\ D {\bf 61}, 043511 (2000)
  [astro-ph/9909480].
  %%CITATION = ASTRO-PH/9909480;%%

%\cite{Kawai:1999pw}
\bibitem{Kawai:1999pw} 
  S.~Kawai and J.~Soda,
  %``Evolution of fluctuations during graceful exit in string cosmology,''
  Phys.\ Lett.\ B {\bf 460}, 41 (1999)
  [gr-qc/9903017].
  %%CITATION = GR-QC/9903017;%%

%\cite{Kawai:1999xn}
\bibitem{Kawai:1999xn} 
  S.~Kawai and J.~Soda,
  %``Structure formation from nonsingular kinetic inflation,''
  gr-qc/9906046.
  %%CITATION = GR-QC/9906046;%%

%\cite{Cartier:2001is}
\bibitem{Cartier:2001is} 
  C.~Cartier, J.~c.~Hwang and E.~J.~Copeland,
  %``Evolution of cosmological perturbations in nonsingular string cosmologies,''
  Phys.\ Rev.\ D {\bf 64}, 103504 (2001)
  [astro-ph/0106197].
  %%CITATION = ASTRO-PH/0106197;%%

%\cite{Kobayashi:2011nu}
\bibitem{Kobayashi:2011nu} 
  T.~Kobayashi, M.~Yamaguchi and J.~Yokoyama,
  %``Generalized G-inflation: Inflation with the most general second-order field equations,''
  Prog.\ Theor.\ Phys.\  {\bf 126}, 511 (2011)
  doi:10.1143/PTP.126.511
  [arXiv:1105.5723 [hep-th]].
  %%CITATION = doi:10.1143/PTP.126.511;%%



%\cite{Planck:2013jfk}
\bibitem{Planck:2013jfk} 
  P.~A.~R.~Ade {\it et al.} [Planck Collaboration],
  %``Planck 2013 results. XXII. Constraints on inflation,''
  Astron.\ Astrophys.\  {\bf 571}, A22 (2014)
  doi:10.1051/0004-6361/201321569
  [arXiv:1303.5082 [astro-ph.CO]].
  %%CITATION = doi:10.1051/0004-6361/201321569;%%

%\cite{Neupane:2007qw}
\bibitem{Neupane:2007qw} 
  I.~P.~Neupane,
  %``Constraints on Gauss-Bonnet cosmologies,''
  arXiv:0711.3234 [hep-th];\\
  %%CITATION = ARXIV:0711.3234;%%
  M.~Satoh, S.~Kanno and J.~Soda,
  %``Circular Polarization of Primordial Gravitational Waves in String-inspired Inflationary Cosmology,''
  Phys.\ Rev.\ D {\bf 77}, 023526 (2008)
  [arXiv:0706.3585 [astro-ph]];\\
  %%CITATION = ARXIV:0706.3585;%%
  M.~Satoh and J.~Soda,
  %``Higher Curvature Corrections to Primordial Fluctuations in Slow-roll Inflation,''
  JCAP {\bf 0809}, 019 (2008)
  [arXiv:0806.4594 [astro-ph]];\\
  %%CITATION = ARXIV:0806.4594;%%
  Z.~K.~Guo and D.~J.~Schwarz,
  %``Power spectra from an inflaton coupled to the Gauss-Bonnet term,''
  Phys.\ Rev.\ D {\bf 80}, 063523 (2009)
  [arXiv:0907.0427 [hep-th]];\\
  %%CITATION = ARXIV:0907.0427;%%
  M.~Satoh,
  %``Slow-roll Inflation with the Gauss-Bonnet and Chern-Simons Corrections,''
  JCAP {\bf 1011}, 024 (2010)
  [arXiv:1008.2724 [astro-ph.CO]];\\
  %%CITATION = ARXIV:1008.2724;%%
  S.~Koh, B.~H.~Lee, W.~Lee and G.~Tumurtushaa,
  %``Observational constraints on slow-roll inflation coupled to a Gauss-Bonnet term,''
  Phys.\ Rev.\ D {\bf 90}, no. 6, 063527 (2014)
  [arXiv:1404.6096 [gr-qc]];\\
  %%CITATION = ARXIV:1404.6096;%%
  A.~De Felice, S.~Tsujikawa, J.~Elliston and R.~Tavakol,
  %``Chaotic inflation in modified gravitational theories,''
  JCAP {\bf 1108}, 021 (2011)
  doi:10.1088/1475-7516/2011/08/021
  [arXiv:1105.4685 [astro-ph.CO]].
  %%CITATION = doi:10.1088/1475-7516/2011/08/021;%%

%\cite{Weinberg:2008hq}
\bibitem{Weinberg:2008hq} 
  S.~Weinberg,
  %``Effective Field Theory for Inflation,''
  Phys.\ Rev.\ D {\bf 77}, 123541 (2008)
  [arXiv:0804.4291 [hep-th]];\\
  %%CITATION = ARXIV:0804.4291;%%
  D.~Baumann, H.~Lee and G.~L.~Pimentel,
  %``High-Scale Inflation and the Tensor Tilt,''
  arXiv:1507.07250 [hep-th].
  %%CITATION = ARXIV:1507.07250;%%

\end{thebibliography}
\end{document}